\documentclass[doublespacing]{elsart}

\usepackage[english]{babel}

\usepackage{amsmath, amssymb, amsfonts}
\usepackage{epsf}
\usepackage{graphicx}
\usepackage{cite}
\usepackage{color,soul}

\begin{document}
	
	\begin{frontmatter}
		
		\title{Electromagnetic dipole, quadrupole moments and parity and time reversal invariance interactions of $\Omega^{\pm}$ baryons in bent and straight crystals}
		
		\author{V.G.~Baryshevsky}
		
		\address{Institute for Nuclear Problems, Belarusian State University,
			Bobruiskaya 11, 220030 Minsk, Belarus, bar@inp.bsu.by, v\_baryshevsky@yahoo.com}
		
		\begin{abstract}
			Experiments with $\Omega^{\pm}$ baryons moving in a crystal at LHC in addition to
			EDM measurement open possibility to study electric quadrupole
			moment, to observe birefringence effect (spin oscillations and
			spin dichroism) for $\Omega^{\pm}$ baryons and to get information
			about constants describing P-odd, T-odd effects at $\Omega^{\pm}$
			baryon interactions with electrons and nuclei
		\end{abstract}
		
		
	\end{frontmatter}

    \section{Introduction}

Recently, the experimental approach was proposed \cite{Phys_1,Phys_2,VG_3,VG_4} to search for the electromagnetic dipole moments (EDM) of charged short-lived heavy baryons and $\tau$-leptons using bent crystal at LHC.
According \cite{VG_x,VG_6,VG_7}, the same approach gives a unique possibility for investigating the P-odd T-even and P-odd T-odd (CP-odd) interactions of short-lived baryons ($\tau$-leptons) with electrons and nuclei.
It was also shown that measuring the polarization vector and the angular distribution of 
charged and neutral particles, scattered by axes (planes) of an unbent (straight) crystal 
enables to obtain restrictions for the EDM value and for magnitudes of constants describing T-odd 
(CP-odd) interactions beyond the standard model.
Possibility to measure the quadrupole moment of $\Omega^\pm$ baryon   
was also demonstrated  \cite{Phys_5,Phys_6}.
In present paper the expression for effective potential energy for $\Omega^\pm$ baryon in a crystal is obtained that enables formulation of equations, which describe evolution of spin of a $\Omega^\pm$ baryon moving in a crystal.
Principal phenomena, which influence on spin evolution and enable investigation of  the quadrupole moment of $\Omega^\pm$ baryon are indicated. They are as follows: birefringence effect and different T-odd,P-odd  interactions of $\Omega^\pm$ baryon with electrons and nuclei.

\section{Spin interactions of relativistic $\Omega^{\pm}$ baryon with crystals}

For high-energy particles moving in intracrystalline electromagnetic fields the particle wavelength $\lambda$ is much smaller as compared to the typical field nonuniformity: $\lambda\ll R_{sh}$, where $R_{sh}$ is the shielding radius.
This fact enables applying the quasi-classical approximation. As a result to study particle spin evolution in the electromagnetic fields inside the crystal one can use
the Thomas-Bargmann-Michael-Telegdi (T-BMT) equations. The T-BMT equation describes spin motion in the rest frame of the particle, wherein spin is described by three component vector $\vec{S}$ \cite{Phys_5,Phys_13,Phys_26}. The term ''particle spin'' here means the expected value of the quantum mechanical spin operator $\hat{\vec{S}}$, hereinafter the symbol marked with a ''hat'' means a quantum mechanical operator.

Modification of the T-BMT equation allowing for the possible existence of non-zero electric dipole moment (EDM) for a particle moving in a crystal were studied in  \cite{Phys_1,Phys_2,VG_3,VG_4,VG_x,VG_6,VG_7,Phys_5,Phys_6}.

Futher description will be done for a relativistic particle, which moves in the non-magnetic crystal. Let us suppose particle Lorentz-factor $\gamma$ to be high: $\gamma \gg 1$. In this case spin motion is described by:
\begin{equation}\label{eq1}
\frac{d\vec{S}}{dt}=[\vec{S}\times\vec{\Omega}_{magn}],
\end{equation}
\begin{equation}\label{eq2}
\vec{\Omega}_{magn}=-\frac{e(g-2)}{2mc}[\vec{\beta}\times\vec{E}],
\end{equation}
where $\vec{S}$ is the spin vector, $t$ is the time in the laboratory frame, $m$ is the mass of the particle, $e$ is its charge, $\vec{\beta}=\frac{\vec{v}}{c}$, where $\vec{v}$ denotes the particle velocity, $\vec{E}$ is the electric field at the point of particle location in the laboratory frame, and $g$ is the gyromagnetic ratio. By definition, the particle magnetic moment is $\mu=(eg\hbar /2mc)S$.\\
If a particle possesses an intrinsic electrical dipole moment, then the additional term, describing spin rotation induced by the EDM, should be added to (\ref{eq1}):
\begin{equation}\label{eq3}
\frac{d\vec{S}_{EDM}}{dt}=[\vec{S}\times\vec{\Omega}_{EDM}],
\end{equation}
\begin{equation}\label{eq4}
\vec{\Omega}_{EDM}=\frac{D}{S \hbar}\left\{\vec{E}-\frac{\gamma}{\gamma+1}\vec{\beta}(\vec{\beta}\vec{E})\right\},
\end{equation}
where $D=ed$ is the electric dipole moment of the particle.\\
As a result, motion of a particle spin due to the magnetic and electric dipole moments can be described by the following equation:
\begin{equation}\label{eq5}
\frac{d\vec{S}}{dt}=[\vec{S}\times\vec{\Omega}_{magn}]+[\vec{S}\times\vec{\Omega}_{EDM}].
\end{equation}

The above equation is not applicable for studies dedicated for EDM search of $\Omega^\pm$ baryon, because its spin $S=\frac{3}{2}>\frac{1}{2}$ and $\Omega^\pm$ baryon can possess quadrupole moment.
This is the reason to generalize (\ref{eq5}) by adding the term, which describes interaction of relativistic particle quadrupole moment with nonuniform electric field of the crystal.
The corresponding equation was obtained in  \cite{Phys_5,12,13} and reads as follows:
\begin{equation}\label{eq6}
\frac{d\hat{S}_i}{dt}=\frac{e}{3\hbar}\varepsilon_{ikl}\varphi_{kn}\hat{Q}_{ln},
\end{equation}
where $\hat{S}_i$ is the operator of the particle spin projection; repeated indices imply summation;
\begin{equation}\label{eq7}
\hat{Q}_{ln}=\frac{3Q}{2S(2S-1)} \left\{\hat{S}_{ln}-\frac{2}{3}S(S+1)\delta_{ln}\right\}
\end{equation}
is the operator of quadrupole moment tensor, Q is the particle quadrupole moment, $\hat{S}_{ln}=\hat{S}_i\hat{S}_n+\hat{S}_n\hat{S}_i$, 
$\varphi_{kn}=\frac{\delta^2\varphi}{\delta x_k\delta x_n}$, 
the electric potential $\varphi$ is evaluated in the point of particle location in the crystal at the instant $t$,
$\varepsilon_{inl}$ is the totally antisymmetric unit tensor.

As a result, the equations, which describes spin evolution for a particle with spin $S\ge 1$ moving in a crystal, can be expressed as follows:
\begin{equation}\label{eq8}
\frac{d\hat{S}_i}{dt}=[\hat{\vec{S}}\times\vec{\Omega}_{magn}]_i+[\hat{\vec{S}}\times\vec{\Omega}_{EDM}]_i+
\frac{e}{3\hbar}\varepsilon_{ikl}\varphi_{kn}\hat{Q}_{ln}.
\end{equation}
The above equation depends on $\hat{Q}_{ln}$. Using (\ref{eq7}) and (\ref{eq8}) one can obtain the equation for $\frac{d\hat{Q}_{ln}}{dt}$ \cite{Phys_5,12,13}.
Note that the spin matrix of dimensionality $(2S+1)(2S+1)$ can be expanded in terms of a complete set of $(2S+1)^2$ matrices, particularly, in terms of a set of polarization operators $\hat{T}_{LM}(S)$, where $0\ll L\ll 2S$, $-L\ll M\ll L$. 
The polarization operator is an irreducible tensor of rank $L$. The maximal rank of $\hat{T}_{LM}(S)$ is $2S$. 
The same matrix can be expanded in terms of a set of Cartesian tensors of maximal rank $2S$, i.e., in terms of a set of products $S_iS_kS_l$...$S_m$ with the maximal number of factors in this product equal to $2S$. These Cartesian tensors are reducible and may be represented as a sum of irreducible tensors.

 Therefore, for the case of interest, namely:  for $\Omega^{\pm}$ baryon with spin $S=\frac{3}{2}$, the number of irreducible tensors is 3 (in case of a deuteron, which spin is $S=1$ the number of irreducible tensors is 2)

This is the reason, why equation (\ref{eq8}) includes combinations of three spin operators  $\hat{S}_i\hat{S}_k\hat{S}_{n}$ along with $\hat{S}_i$ and $\hat{Q}_{ln}$.
To get the equation, which in general case describes spin evolution for  $\Omega^{\pm}$ baryon, let us use quantum mechanics guidlines.
 The operator equation describing the time evolution of an arbitrary quantum mechanical operator can be written as follows:
\begin{equation}\label{eq9}
\frac{d\hat{S}}{dt}=\frac{i}{\hbar} [\hat{H}\hat{S}],
\end{equation}
where $\hat{H}$ is the Hamiltonian.\\
\\
Upon calculating the commutation relation of $\hat{H}$ and $\hat{S}$, we find out that due to the presence of the interactions particle with crystal, the commutation relation for a particle with spin $S\ge 1$, in contrast to that for a particle with spin $S=\frac{1}{2}$, will contain the product of the components of operator $\hat{S}_i$ in addiction to operator $\hat{S}$. When the particle spin is $S=\frac{3}{2}$, the products of the form $\hat{S}_i\hat{S}_k$ and $\hat{S}_i\hat{S}_k\hat{S}_{n}$ will appear.\\
In a similar way, in writing the Heisenberg equation for operator $\hat{S}_i\hat{S}_k$(and for operator $\hat{S}_i\hat{S}_k\hat{S}_{n}$) we shall obtain the two equations that include $\hat{S}_i\hat{S}_k$, $\hat{S}_i\hat{S}_k\hat{S}_{n}$ and $\hat{S}_i$. As a result, we obtain the system of three connected equations for operators $\hat{S}_i$, $\hat{S}_i\hat{S}_k$ and $\hat{S}_i\hat{S}_k\hat{S}_{n}$.

Therefore, in case under consideration, the operator equation describing the time evolution of a spin operator can be written as \cite{Phys_5}:
\begin{equation}\label{eq10}
\frac{d\hat{S}}{dt}=\frac{i}{\hbar} [\hat{U}\hat{S}],
\end{equation}
where $\hat{U}$ is the spin-dependent effective potential energy for a particle moving in a crystal.
According to \cite{VG_6,VG_7,Phys_5} the spin-dependent effective potential energy of
a particle moving in a crystal is a periodic function of particle coordinates:
\begin{equation}\label{eq11}
\hat{U}(\vec{r})=\sum_{\vec{\tau}}\hat{U}(\vec{\tau})e^{i \vec{\tau} \vec{r}},
\end{equation}
where $\vec{\tau}$ is the reciprocal lattice vector of the crystal,
\begin{equation}\label{eq12}
\hat{U}(\vec{\tau})=\frac{1}{V}\sum_{j}\hat{U}_j (\vec{\tau})e^{i\vec{\tau}\vec{r}_j}.
\end{equation}
Here, $V$ is the volume of the crystal elementary cell, $\vec{r}_j$ is the coordinate for the atom (nucleus) of type j in the crystal elementary cell, and:
\begin{equation}\label{eq13}
\hat{U}_j(\vec{\tau})=-\frac{2\pi\hbar^2}{m\gamma} \hat{F}_j(\vec{\tau}),
\end{equation}
where $m$ - mass of the particle, $\gamma$ is the Lorentz factor. $\hat{F}_j(\vec{\tau})=\hat{F}_j(\vec{q}=\vec{\tau})$ is the amplitude of elastic coherent scattering by the atom (nucleus) of type j, $\vec{q}=\vec{k}'-\vec{k}, \vec{k}'$ is a wave vector of the scattered particle, $\vec{k}$ is a wave vector of the incident particle.

Thus, to use equations (\ref{eq10},\ref{eq11}) for defining the system of equations, which describes evolution of spin of a particle moving in a crystal, one should find the expression for amplitude of elastic coherent scattering  $\hat{F}(\vec{q})$.

 The expression for $\hat{F}(\vec{q})$ for baryons with spin $\frac{1}{2}$ in presence of P-violating and T-violating interactions is obtained in  \cite{VG_6}.

For particles with spin $\frac{3}{2}$ the amplitude $\hat{F}(\vec{q})$  includes additional terms proportional to $S_iS_l$ and $S_iS_lS_n$.
These Cartesian tensors are reducible and may be represented as a sum of irreducible tensors.

The most general form of amplitude of elastic cogerent scattering $\hat{F}(\vec{q})$ can be obtained  if consider the fact that $\hat{F}(\vec{q})$ should be scalar with respect to rotations.

The scalar combinations one should form using $S_i, \vec{k}$ and $\vec{k}'$, where $\vec{k}$ is the wave vector of the particle before collision (particle momentum $\vec{p}=\hbar\vec{k}$), $\vec{k}'$ is the wave vector of the scattered particle (particle momentum $\vec{p}'=\hbar\vec{k}'$).

Since scattering is supposed to be elastic, then $|\vec{k}|=|\vec{k}'|$.
The following momentum combinations is convenient to be introduced $[\vec{k}'\times\vec{k}], \vec{k}'+\vec{k}$ and $\vec{k}'-\vec{k}$ along with the corresponding unitary vectors
 \cite{VG_7}:
\begin{center}
    $\vec{N}=\frac{[\vec{k}'\times\vec{k}]}{|\vec{k}'\times\vec{k}|}, \vec{N}_w=\frac{\vec{k}'+\vec{k}}{|\vec{k}'+\vec{k}|}, N_T=\frac{\vec{k}'+\vec{k}}{|\vec{k}'-\vec{k}|}$.
\end{center}
Considering the above one can write the following expression for  $\hat{F}(\vec{q})$:
\begin{multline}\label{eq14}
\hat{F}(\vec{q})=A(\vec{q})+B(\vec{q})\hat{\vec{S}}\vec{N}+
B_w(\vec{q})\hat{\vec{S}}\vec{N}_w+B_T(\vec{q})\hat{\vec{S}}\vec{N}_T+
D\hat{S}_i\hat{S}_l N_{wi}N_{wl}+\\
+D_1\hat{S}_i\hat{S}_lN_{Ti}N_{Tl}+D_2\hat{S}_i\hat{S}_lN_iN_l+D_3\hat{S}_i\hat{S}_lN_iN_{wi}+D_4S_iS_lN_iN_{Tl}+\\
+D_5S_iS_lN_{wi}N_{Tl}+G\hat{S}_i\hat{S}_l\hat{S}_nN_iN_lN_n+...
\end{multline}
here omission points denote large group of summands, which can be formed as a result of multiplication of tensors $\hat{S}_i\hat{S}_l\hat{S}_n$ with combinations of unitary vectors  $\vec{N},\vec{N}_w and \vec{N}_T$; repeated indices imply summation.

Substitution of $\hat{F}(\vec{q})$ into expression for the effective potential energy  $\hat{U}(\vec{r})$ allows to obtain implicit expression for spin-dependent potential energy $\hat{U}(\vec{r})$ and, therefore, using (\ref{eq9},\ref{eq10}), the equation, which described spin evolution for a particle in a crystal.

Let us analyze expression (\ref{eq14}) in details. The first group of terms, which comprises spin operator, was studied in \cite{VG_x,VG_6,VG_7,Phys_5,Phys_6}.
Let us now consider the terms proportional to the productions of spin operators $\hat{S}_i \hat{S}_l$.
Note that for forward scattering (i.e. for $\vec{q}=0$) the scattering amplitude $\vec{F}(\vec{q}=0)$ the terms comprising product of spin operators $\hat{S}_i\hat{S}_l$ are all equal to zero, besides that one, which is proportional to $D$.
This term describes birefringence effect for particles with spin  $S \ge 1$ \cite{Phys_5}.
The birefringence effect was experimentally observed for deuteron (see in details in \cite{Phys_5}).
Experimental study of birefringence effect for high-energy $\Omega^\pm$ baryons enable study of quark-quark scattering \cite{Phys_5}.

When $\vec{q}\neq 0$, the term  comprising $D_1$ describes effect of spin oscillation and rotation for a baryon channeled in a crystal.
The above effect is caused by particle EDM \cite{VG_6,VG_7} (see also (\ref{eq6})).
The term proportional to $D_2$ describes contribution to oscillations, which is due to magnetic and spin-orbit nuclear interactions of a particle channeled in a crystal.
$D_3$-proportional term is the P-odd, T-even summand, those comprising $D_4$ and $D_5$ are responsible for T-odd P-odd and T-odd P-even interactions.
Among the terms proportional to $\hat{S}_i\hat{S}_l\hat{S}_n$ T-odd P-even terms are also present.

Therefore, study of spin oscillations and spin dichroism for $\Omega^\pm$ baryons moving in a crystal enables investigating different  P-odd and T-odd interactions along with EDM measuring.

As a first step one could consider the experiment for measuring  the quadrupole moment and birefringence effect for $\Omega^\pm$ baryon.

It is good to mention that birefringence effect reveals for $\Omega^\pm$ baryon moving in either crystal or chaotic media.

When planning the experiments one should also realize that for $\Omega^\pm$ baryon moving into the target, vector-to-tensor (tensor-to-vector) polarization conversion will present  along with spin oscillation (rotation).

This fact allows to choose the optimal conditions for effect observing.

Thus, study of $\Omega^\pm$ baryon motion in a crystal makes it possible to study various effects, namely:
\begin{enumerate}
    \item Spin rotation in a bent crystal caused by baryon anomalous magnetic moment;
    \item Spin rotation, caused by baryon EDM and other T-odd effects;
    \item Spin rotation and oscillation, vector-to-tensor (tensor-to-vector) polarization conversion caused by quadrupole moment, birefringence and P-odd, T-odd effects in both straight and bent crystals.
\end{enumerate}
All the above promises to obtain new information about $\Omega^\pm$ baryon properties at experiments at LHC and future colliders. In particular, quadrupole moment  and birefringence effect can be measured. Evaluation of P-odd, T-odd and P-T-odd effects can be obtained.

    \section{Conclusion}
In present paper the effective potential energy for a $\Omega^\pm$ baryon in a
crystal is expressed that enables formulation of equations, which describe
evolution of spin of a $\Omega^\pm$ baryon moving in a crystal. Principal phenomena,
which influence on spin evolution and enable investigation of the quadrupole
moment of $\Omega^\pm$ baryon are indicated. Experiments with $\Omega^\pm$ baryons moving in a crystal at LHC in addition to EDM measurement open possibility to study electric quadrupole moment, to observe birefringence effect (spin oscillations and spin dichroism) for $\Omega^\pm$ baryons and to get information about constants describing P-odd, T-odd effects at $\Omega^\pm$ baryon interactions with electrons and nuclei

\end{document}